\documentclass[a4paper]{jpconf}
\usepackage{graphicx}
\begin{document}
\title{Status of the ANAIS Dark Matter Project at the Canfranc Underground Laboratory}

\author{J.~Amar\'e, S.~Cebri\'an, C.~Cuesta\footnote{Present address: Center for Experimental Nuclear Physics and Astrophysics and Department of Physics, University of Washington, WA, US}, E.~Garc\'{\i}a, M.~Mart\'{\i}nez\footnote{Present address: Universit\`a degli Studi di Roma "La Sapienza", Roma, Italy}, M.A.~Oliv\'an, Y.~Ortigoza, A.~Ortiz de Sol\'orzano, C.~Pobes\footnote{Present address: Instituto de Ciencia de Materiales de Arag\'on, Universidad de Zaragoza - CSIC, Zaragoza, Spain} , J.~Puimed\'on, M.L.~Sarsa, J.A.~Villar and P.~Villar}

\address{Laboratorio de F\'{\i}sica Nuclear y Astropart\'{\i}culas, Universidad de Zaragoza, C/~Pedro Cerbuna 12, 50009 Zaragoza, SPAIN\\
Laboratorio Subterr\'aneo de Canfranc, Paseo de los Ayerbe s.n., 22880 Canfranc Estaci\'on, Huesca, SPAIN}

\ead{mlsarsa@unizar.es}

\begin{abstract}
The ANAIS (Annual modulation with NaI(Tl) Scintillators) experiment aims at the confirmation of the DAMA/LIBRA signal using the same target and technique at the Canfranc Underground Laboratory (LSC). Along 2016, 112.5\,kg of ultra pure NaI(Tl) crystals will be installed at LSC in a 3x3 modules matrix configuration. The ANAIS-25 and ANAIS-37 set-ups have been taking data at the LSC testing the detector performance, the DAQ and analysis systems, and assessing the background. Main results coming from both set-ups will be summarized in this paper, focusing on the excellent detector performance and background understanding. Prospects for the experiment will be also briefly revised.
\end{abstract}

\section{Introduction}
\label{sec:intro}

ANAIS project aims at the study of the annual modulation signal attributed to galactic dark matter particles\,\cite{annual_modulation} with NaI(Tl) detectors at the Canfranc Underground Laboratory (LSC), in Spain. For many years the DAMA/LIBRA experiment, at the Laboratori Nazionali del Gran Sasso, in Italy, has reported the presence of an annual modulation in the detection rate compatible with that expected for a dark matter signal with a very high statistical significance\,\cite{DAMA,LIBRA}. Other very sensitive experiments have put stringent limits on the dark matter particles properties, constraining the compatibility scenarios, but comparison is model-dependent. ANAIS, using the same target than DAMA/LIBRA, would enable a model independent test of the DAMA/LIBRA signal. To achieve this goal, ANAIS plans to install 112.5\,kg of NaI(Tl) detectors at LSC along 2016 in a 3x3 modules configuration.

Several prototypes have been operating at the LSC in the last years\,\cite{ANAIS_prototypes, ANAISbkg, ANAIS_RICAP, ANAIS250}, enabling a robust knowledge of the detector performance and background contributions\,\cite{ANAISbkg, ANAISom, ANAISquartz, ANAIS_40K, ANAIS_cosmogenics}. In the following, it will be reported on the main results derived from the ANAIS-25 and ANAIS-37 set-ups: firstly, we will briefly describe the experimental set-up (Section\,\ref{sec:setup}); then, it will be highlighted the excellent light collection and triggering at very low energy (Section\,\ref{sec:detperformance}); the filtering of photomultiplier tubes (PMT) events and other anomalous scintillation (Section\,\ref{sec:filtering}); and the remarkable background understanding (Section\,\ref{sec:bck}). Finally, prospects for the experiment will be revised (Section\,\ref{sec:prospects}).

\section{Experimental set-up}
\label{sec:setup}

\begin{figure}[h!]
\includegraphics[width=0.7\textwidth]{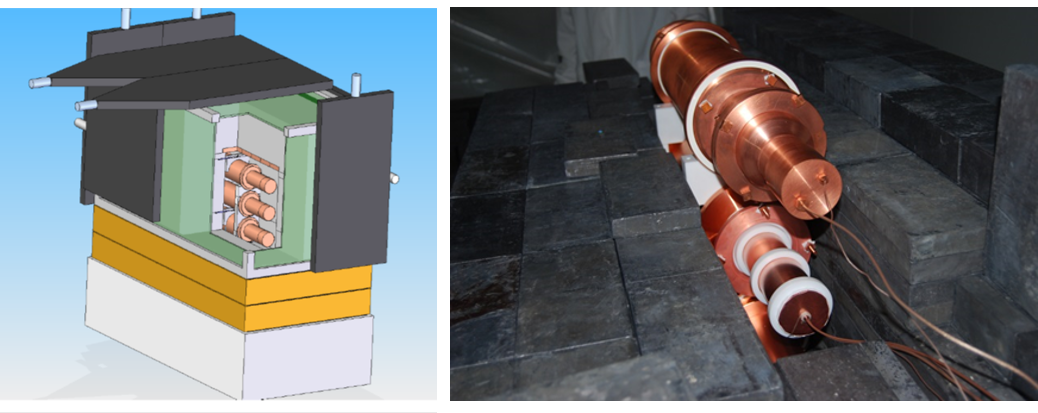} \hspace{1pc}%
\begin{minipage}[b]{9pc}\caption{ANAIS-37 set-up at LSC.}
\label{fig:ANAIS-37}
\end{minipage}
\end{figure}

The ANAIS-25 set-up consisted of two cylindrical 12.5\,kg mass NaI(Tl) crystals grown by Alpha Spectra\,\footnote{Alpha Spectra Inc., Grand Junction, Colorado, US. http://www.alphaspectra.com/} (AS), and housed in OFHC (Oxygen-Free High thermal Conductivity) copper with two synthetic quartz windows allowing for the coupling of the PMTs in a second step at the LSC clean room. We will refer to both modules as D0 and D1 in the following. An aluminized Mylar window allowed for calibration at very low energy. Hamamatsu R12669SEL2 PMTs have been used in most of the results presented in this paper. Shielding consisted of 10\,cm archaeological lead, 20\,cm low activity lead, PVC box tightly closed and continuously flushed with boil-off nitrogen gas, and active plastic scintillator vetoes placed on top of the shielding.
The ANAIS-37 set-up refers to the enlargment of ANAIS-25 in order to include a new improved module (D2), 12.5\,kg mass, placed in between D0 and D1. Same PMT model and shielding described above were used (see figure~\ref{fig:ANAIS-37}).

\section{ANAIS detectors performance}
\label{sec:detperformance}

Total light collected per unit of energy deposited in different NaI(Tl) crystals (see table~\ref{tab:yield}) has been calculated with the mean pulse area of events corresponding to 22.6\,keV from a $^{109}$Cd source and the mean pulse area of single photoelectrons (phe), identified in pulses with a few number of them and averaged to build the Single Electron Response (SER). An excellent light yield in all the AS modules can be reported, pointing at the possibility of further reduction of the ANAIS threshold, already demonstrated to be at 2\,keVee level in the ANAIS-0 prototype\,\cite{ANAISfiltrado}, consisting of a Saint-Gobain crystal, 9.6\,kg mass, similar to those used by DAMA-LIBRA.

\begin{table}[ht]
\begin{center}
\caption{Results for the effective amount of light collected per unit of energy deposited in different NaI(Tl) crystals. Results using two different PMT models are shown, if available.}
{\begin{tabular}{@{}lll@{}}
\br
PMT model  & NaI(Tl) crystal & Light Collected\\
 &  & (phe/keV)\\		
\mr
Ham R12669SEL2 / Ham R11065SEL &  ANAIS-0 &  \hphantom{0}$7.4 \pm 0.1$ / \hphantom{0}$5.3\pm 0.1 $ \\
Ham R12669SEL2 & D0 & $15.6 \pm 0.2$  \\
Ham R12669SEL2 / Ham R11065SEL & D1 &  $15.2 \pm 0.1$ /  $12.6 \pm 0.1 $  \\
Ham R12669SEL2 & D2 &  $16.3 \pm 0.6$ \\
\br
\end{tabular}
\label{tab:yield}}
\end{center}
\end{table}

\begin{figure}[h!]
\includegraphics[width=1.\textwidth]{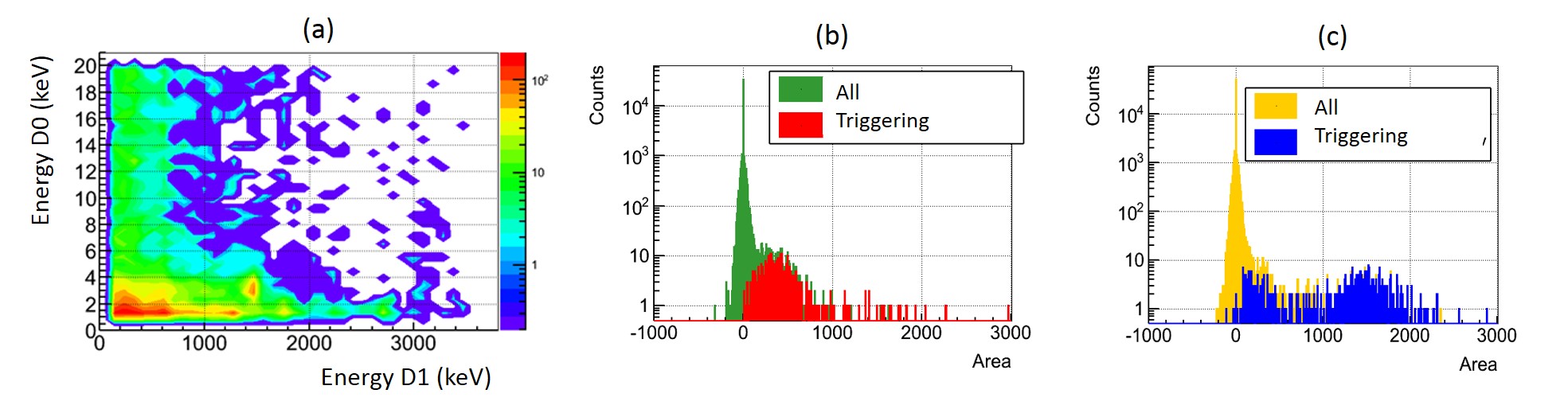}
\caption{a) Coincidence plot between low energy depositions in D0 and high energy in D1 in ANAIS-25 set-up. b) Low energy events in D0 in coincidence with 1275\,keV energy deposition in D1 are shown in green, those triggering in red; they correspond to 0.9\,keVee. c) Low energy events in D0 in coincidence with 1461 keV energy deposition in D1 are shown in yellow, those triggering in blue; they correspond to 3.2 and 0.3\,keVee.}
\label{fig:trigger}
\end{figure}

Triggering is done at phe level in both PMTs in AND logical mode (200\,ns window) for every module and in OR logical mode between modules. Energy is estimated through the area of the digitized pulse (1.25\,$\mu$s window at 2\,Gs/s sampling rate). Triggering has been checked to be effectively done down to 1\,keVee by looking at the coincidences between high energy gammas of $^{40}$K and $^{22}$Na in one detector and very low energy depositions in a second one, at 3.2 and 0.9\,keVee, respectively (see figure\,\ref{fig:trigger}). Both lines are used for energy calibration of the detectors combined with periodic external calibrations using a $^{109}$Cd source.

\section{Filtering of PMT-origin events and other anomalous scintillation}
\label{sec:filtering}

\begin{figure}[h!]
\includegraphics[width=.65\textwidth]{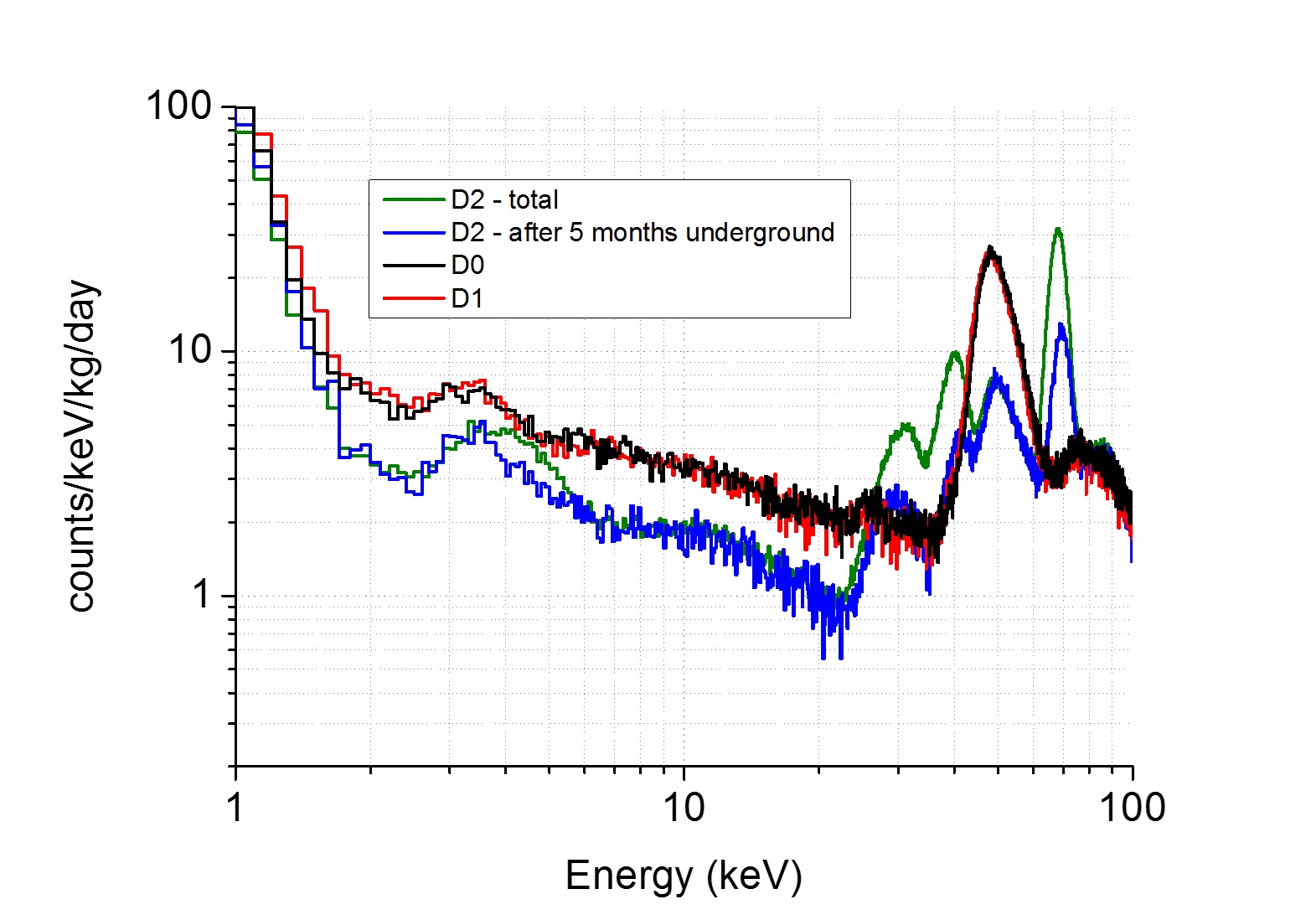}
\begin{minipage}[b]{12pc} \caption{Background at low energy for D0, D1 and D2 modules in the ANAIS-37 set-up. In the case of D2 background level is shown for the first six months underground (green line) and for only the last month, after being five months underground (blue line).}
\label{fig:bkgLE}
\end{minipage}
\end{figure}

Very strong tools to discriminate scintillation events stemming from the NaI(Tl) bulk crystal from other kind of spurious events, having mainly the origin in the PMTs, are being developed. Background below 10\,keVee in NaI(Tl) detectors is strongly dominated by them. ANAIS applies filtering protocols and evaluates the corresponding acceptance efficiencies by studying populations of events at very low energy attributable to bulk NaI(Tl) scintillation, both, from $^{40}$K and $^{22}$Na identified coincidence events, and $^{109}$Cd calibration events. We have checked that very few of those anomalous events are passing the filtering by measuring with a blank module (without crystal) and applying the same data analysis procedures. Low energy background spectra for D0, D1 and D2 modules, after filtering and correcting by the corresponding efficiency, are shown in figure\,\,\ref{fig:bkgLE}. However, our efficiencies for acceptance of bulk events seem to be underestimated between 1 and 2\,keVee and we are working on a better estimate.

\section{Background understanding}
\label{sec:bck}

A successful background model was developed for ANAIS-0 prototype\,\cite{ANAISbkg}, serving as starting point for the building of the background model of the AS modules. Main issues affecting backgrounds in the dark matter region of interest (ROI) in previous prototypes were the internal contaminations in potassium and $^{210}$Pb. The main goal of the ANAIS-25 set-up was to determine the potassium content of D0 and D1 modules. For that, the study of the coincidences between the two modules was used and a result for the $^{40}$K activity of 1.25\,$\pm$\,0.11\,mBq/kg (equivalent to 41.7\,$\pm$\,3.7\,ppb natural potassium), one order of magnitude smaller than that of ANAIS-0, was derived \cite{ANAIS250, ANAIS_40K}. Results of ANAIS-37 set-up have provided more information on the $^{210}$Pb and potassium content of D0, D1 and D2 modules (see table~\ref{tab:contaminants}) and very promising upper limits on $^{238}$U and $^{232}$Th levels, being work still in progress.

\begin{table}[ht]
\begin{center}
\caption{Results derived from ANAIS-37 data for the radiopurity of bulk NaI(Tl) in AS modules.}
{\begin{tabular}{@{}lll@{}}
\br
Isotope & NaI(Tl) module &  Activity\\
 &  & (mBq/kg)\\		
\mr
 & D0 &   $ 1.4 \pm 0.2$  \\
$^{40}$K &  D1  &  $1.1\pm 0.2 $ \\
  & D2  & $1.1 \pm 0.2$  \\
\mr
 $^{210}$Pb & D0 / D1&  $3.15 \pm 0.10$  \\
 &D2  & $0.70 \pm 0.10$ \\
\br
\end{tabular}
\label{tab:contaminants}}
\end{center}
\end{table}

Contaminations in the NaI crystal bulk, as well as those from PMTs and other detector components determined by HPGe Spectrometry have been included in the simulation using Geant4 package and following a similar approach to that presented in\,\cite{ANAISbkg}. In particular, in the crystal bulk have been simulated: $^{40}$K, $^{210}$Pb, cosmogenic isotopes of short lifetime as well as $^{129}$I, $^3$H, and the isotopes from the $^{238}$U and $^{232}$Th chains at the level of previous prototypes. Figure\,\ref{fig:bkgsim} shows a preliminary estimate of the different background contributions up to 100\,keVee in D2.

\section{Sensitivity prospects}
\label{sec:prospects}

Although cosmogenically activated isotopes contribution\,\cite{ANAIS_RICAP, ANAIS_cosmogenics} can be clearly identified in D2 data (see figure\,\ref{fig:bkgLE}) because it is still decaying, the background level at low energy has significantly improved with respect to previous detectors and further improvement is expected in next prototype (D3), presently under construction at AS. Moreover, a large rejection of the $^{40}$K contribution in the ROI could be achieved by vetoing those events by the detection of the associated high energy gamma in a liquid scintillator surrounding the detectors. We are working in the evaluation of the possibilities of such a system and its feasibility in the context of the ANAIS project. 

\begin{figure}[h]
\begin{minipage}{18pc}
\includegraphics[width=18pc]{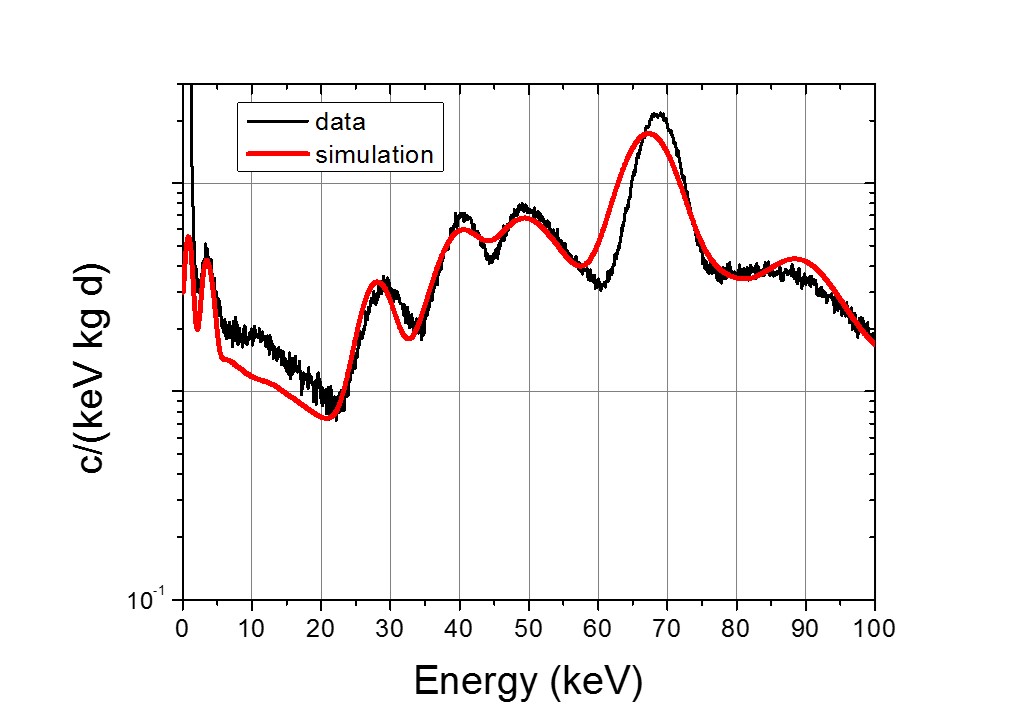}
\caption{\label{fig:bkgsim}Comparison of the D2 low energy background (black line) with the still under construction background model (solid line).}
\end{minipage}\hspace{2pc}%
\begin{minipage}{18pc}
\includegraphics[width=18pc]{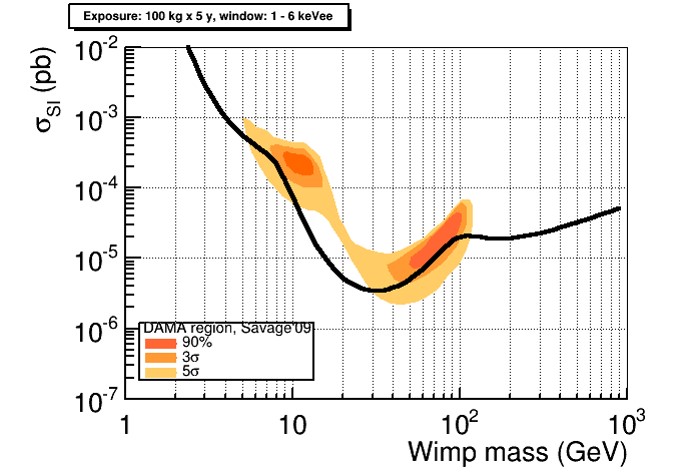}
\caption{\label{fig:prospects}Prospects for ANAIS with 100\,kg and 5\,years in a conservative background scenario.}
\end{minipage}
\end{figure}

Discovery potential of ANAIS3x3 is high, as shown in figure\,\ref{fig:prospects}, even in a very conservative background scenario: assuming 100\,kg of NaI(Tl), 5\,years of data taking and the background already achieved, 90\%\,C.L. positive signal in 90\%\, of the carried out experiments would be obtained for most of the DAMA-LIBRA singled-out dark matter parameter space\,\cite{Savage}.

\section*{Acknowledgments}
This work has been supported by the Spanish Ministerio de Econom\'{\i}a y Competitividad and the European Regional Development Fund (MINECO-FEDER) (FPA2011-23749 and FPA2014-55986-P), the Consolider-Ingenio 2010 Programme under grants MULTIDARK CSD2009- 00064 and CPAN CSD2007-00042, and the Gobierno de Arag\'{o}n and the European Social Fund (Group in Nuclear and Astroparticle Physics). P.\,Villar is supported by the MINECO Subprograma de Formaci\'{o}n de Personal Investigador. We also acknowledge LSC and GIFNA staff for their support.

\end{document}